%Paper: 9202008
%From: Petr Horava <horava@curie.uchicago.edu>
%Date: Mon, 3 Feb 92 16:42:30 CST

\input phyzzx
\overfullrule=0pt
\tolerance=5000
\twelvepoint
%%%%%%%%%%%%%%%%%%%%%%%%%%%%%%%%%%%%%%%%%%%%%%%%%%%%%%%%%%%%%%%%%%%%%%%%%%%
\def\npb#1#2#3{{\it Nucl.\ Phys.} {\bf B#1} (19#2) #3}
\def\plb#1#2#3{{\it Phys.\ Lett.} {\bf #1B} (19#2) #3}
\def\mpl#1#2#3{{\it Mod.\ Phys.\ Lett.} {\bf A#1} (19#2) #3}
\def\prl#1#2#3{{\it Phys.\ Rev.\ Lett.} {\bf #1} (19#2) #3}

\def\cmp#1#2#3{{\it Commun.\ Math.\ Phys.} {\bf #1} (19#2) #3}
\def\ijmp#1#2#3{{\it Int.\ J. Mod.\ Phys.} {\bf A#1} (19#2) #3}
\def\frac#1#2{{#1 \over #2}}

\def\p{\partial}

\def\intz{\int_\Sigma d^2z\;}
     \def\CL{{\cal L}}     \def\CR{{\cal R}}
\def\CG{{\cal G}}     \def\CO{{\cal O}}     \def\CT{{\cal T}}
\def\CJ{{\cal J}}     \def\CQ{{\cal Q}}     

\def\BQ{{\bf Q}}
\def\im{{\rm Im}\,}   
\def\re{{\rm Re}\,}   

%

%%%%%%%%%%%%%%%%%%%%%%%%%%%%%%%%%%%%%%%%%%%%%%%%%%%%%%%%%%%%%%%%%%%%%%%%%%%
\pubnum{EFI-92-06}
\date{January 1992}
\titlepage
%%%%%%%%%%%%%%%%%%%%%%%%%%%%%%%%%%%%%%%%%%%%%%%%%%%%%%%%%%%%%%%%%%%%%%%%%%%
\title{Two Dimensional String Theory and the Topological Torus}
%%%%%%%%%%%%%%%%%%%%%%%%%%%%%%%%%%%%%%%%%%%%%%%%%%%%%%%%%%%%%%%%%%%%%%%%%%%
\vglue-.25in
\author{Petr Ho\v{r}ava\foot{e-mail addresses: horava@yukawa.uchicago.edu
or horava@curie.uchicago.edu}\foot{Robert R. McCormick Fellow;
research also supported by the NSF under Grant No.\ PHY90-00386;
the DOE under Grant No.\ DEFG02-90ER40560;
the Czechoslovak Chart 77 Foundation;
and the \v{C}SAV under Grant No.\ 91-11045.}}
\medskip
\address{\centerline{Enrico Fermi Institute}
\centerline{University of Chicago}
\centerline{5640 South Ellis Avenue}
\centerline{Chicago, IL 60637, USA}}
\bigskip
\abstract{We analyze topological string theory on a two dimensional torus,
focusing on symmetries in the matter sector.  Even before coupling to gravity,
the topological torus has an infinite number of point-like physical
observables, which give rise via the BRST descent equations to an infinite
symmetry algebra of the model.  The point-like observables of ghost
number zero form a topological ground ring, whose generators span a spacetime
manifold;  the symmetry algebra represents all (ground ring valued)
diffeomorphisms of the spacetime.  At nonzero ghost numbers, the
topological ground ring is extended to a superring, the spacetime manifold
becomes a supermanifold, and the symmetry algebra preserves a symplectic form
on it.  In a decompactified limit of cylindrical target topology,
we find a nilpotent charge which behaves like a spacetime topological BRST
operator.  After coupling to topological gravity, this model might represent
a topological phase of $c=1$ string theory.  We also point out some analogies
to two dimensional superstrings with the chiral GSO projection, and to
string theory with $c=-2$.}

\endpage
%%%%%%%%%%%%%%%%%%%%%%%%%%%%%%%%%%%%%%%%%%%%%%%%%%%%%%%%%%%%%%%%%%%%%%%%%%%
\REF\wtopgr{E. Witten, \npb{340}{90}{281};
{\it Surveys in Diff.\ Geom.} {\bf 1} (1991) 243;
and `On the Kontsevich Model and Other Models of Two Dimensional Gravity,'
IAS preprint IASSNS-HEP-91/24 (June 1991)}
\REF\distler{J. Distler, \npb{342}{90}{523}}
\REF\dijkw{R. Dijkgraaf and E. Witten, \npb{342}{90}{486}}
\REF\dvv{E. Verlinde and H. Verlinde, \npb{348}{91}{457};
R. Dijkgraaf, E. Verlinde and H. Verlinde,' \npb{348}{91}{435};
\npb{352}{91}{59}; and in:  `String Theory
and Quantum Gravity,' Proceedings of the Trieste Spring School, April 1990
(World Scientific, Singapore, 1991)}
\REF\kekeli{K. Li, \npb{354}{91}{711}; \npb{354}{91}{725}}
\REF\wgauged{E. Witten, `The $N$ Matrix Model And Gauged WZW Models,'
IAS preprint IASSNS-HEP-91/26 (June 1991)}
\REF\dijk{R. Dijkgraaf, `Intersection Theory, Integrable Hierarchies
and Topological Field Theory,' IAS preprint IASSNS-HEP-91/91 (December 1991)}
\REF\klebrev{I.R. Klebanov, `String Theory in Two Dimensions,' Princeton
preprint PUPT-1271 (July 1991)}
\REF\kutarev{D. Kutasov, `Some Properties of (Non)
Critical Strings,' Princeton preprint PUPT-1277 (September 1991)}
\REF\wsurp{E. Witten, in: `Strings '90,' Proceedings of the Superstring
Workshop at Texas A\& M, March 1990, eds.: R. Arnowitt et al.\ (World
Scientific, Singapore, 1991)}
\REF\polyakov{A.M. Polyakov, \mpl{6}{91}{635};
`Singular States in 2D Quantum Gravity,' Princeton preprint PUPT-1289
(September 1991)}
\REF\oldwitten{E. Witten, \cmp{117}{88}{353}; \cmp{118}{88}{411}}
\REF\wground{E. Witten, `Ground Ring of Two Dimensional String Theory,'
IAS preprint IASSNS-HEP-91/51 (August 1991)}
\REF\klebpol{I.R. Klebanov and A.M. Polyakov, `Interaction of Discrete States
in Two-Dimensional String Theory,' Princeton preprint PUPT-1281 (September
1991); \mpl{6}{91}{3273}}
\REF\miaoli{M. Li, `Correlators of Special States in c=1 Liouville Theory,'
Santa Barbara preprint UCSBTH-91-47 (October 1991)}
\REF\rutground{D. Kutasov, E. Martinec and N. Seiberg, `Ground Rings and Their
Modules in 2D Gravity with $c\leq 1$ Matter,' Princeton \&\ Rutgers
preprint PUPT-1293 \&\ RU-91-49 (November 1991)}
\REF\cernground{P. Bouwknegt, J. McCarthy and K. Pilch, `Ground Ring for the
2D NSR String,' Brandeis \&\ CERN \&\ USC preprint
BRX TH-329 \&\ CERN-TH.6346/91 \&\ USC-91/38 (December 1991)}
\REF\klebward{I.R. Klebanov, `Ward Identities in Two-Dimensional String
Theory,' Princeton preprint PUPT-1302 (December 1991)}
\REF\discrstates{D.J. Gross, I.R. Klebanov and M.J. Newman, \npb{350}{90}{621};
D.J. Gross and I.R. Klebanov, \npb{352}{91}{671}; \npb{359}{91}{3}}
\REF\lizu{B. Lian and G. Zuckerman, \plb{254}{91}{417}; \cmp{135}{91}547;
\plb{266}{91}{21}}
\REF\danielss{U.H. Danielsson and D.J. Gross, \npb{366}{91}{3};
U.H. Danielsson, `Symmetries and Special States in Two Dimensional String
Theory,' Princeton preprint PUPT-1301 (December 1991)}
\REF\ajwinfty{J. Avan and A. Jevicki, \plb{266}{91}{35}; \plb{272}{91}{17};
`String Field Actions from $W_\infty$,' Brown preprint BROWN-HET-839
(October 1991)}
\REF\polchinski{D. Minic, J. Polchinski and Z. Yang, `Translation-Invariant
Backgrounds in 1+1 Dimensional String Theory,' Texas preprint UTTG-16-91}
\REF\mss{G. Moore, N. Seiberg and M. Staudacher, \npb{362}{91}{665}}
\REF\ms{G. Moore and N. Seiberg, `From Loops to Fields in 2D Quantum Gravity,'
Rutgers preprint RU-91-29 (July 1991)}
\REF\conewinf{S.R. Das, A. Dhar, G. Mandal and S.R. Wadia, `Gauge Theory
Formulation of the $c=1$ Matrix Model: Symmetries and Discrete States,'
ETH \&\ IAS \&\ Tata preprint ETH-TH-91/30 \&\ IASSNS-HEP-91/52 \&\
TIFR-TH-91/44 (September 1991); `Bosonization of Nonrelativistic Fermions
and $W$-infinity Algebra,' IAS \&\ Tata preprint IASSNS-HEP-91/72 \&\
TIFR-TH-91/51 (November 1991); `$W$-infinity Ward Identities and Correlation
Functions in the $c=1$ Matrix Model,' IAS and Tata preprint IASSNS-HEP-91/79
\&\ TIFR-TH-91/57 (December 1991)}
\REF\wtoporb{E. Witten, \prl{61}{88}{670}}
\REF\lgtorus{E. Verlinde and N.P. Warner, \plb{269}{91}{96}}
\REF\distvafa{J. Distler and C. Vafa, \mpl{6}{91}{259}; `The Penner Model and
D=1 String Theory,' Princeton preprint PUPT-1212, to appear in the
Proceedings of the Carg\`{e}se Workshop on `Random Surfaces, Quantum Gravity
and Strings,' May 1990}
\REF\japan{Y. Kitazawa, `Puncture Operator
in $c=1$ Liouville Gravity,' Tokyo preprint TIT/HEP-181 (November 1991);
N. Sakai and Y. Tanii, `Operator Product Expansion and Topological
States in $c=1$ Matter Coupled to 2-D Gravity,' Saitama \&\ Tokyo preprint
STUPP-91-22 \&\ TIT/HEP-179 (November 1991)}
\REF\highsym{D.J. Gross, \prl{60}{88}{1229}; E. Witten, {\it Phil.\ Trans.\
Roy.\ Soc.} {\bf A329} (1989) 345}
\REF\wcohom{E. Witten, \ijmp{6}{91}{2775}}
\REF\polch{J. Polchinski, \npb{362}{91}{125}}
\REF\minustwo{I.R. Klebanov and R.B. Wilkinson, \plb{251}{90}{379};
{\it Nucl.\ Phys.} {\bf B} (1991);
J.D. Edwards and I.R. Klebanov, `Macroscopic Boundaries and
the Wave Function of the Universe in the $c=-2$ Matrix Model,' Princeton
preprint PUPT-1250 (March 1991)}
\REF\difrakut{P. Di Francesco and D. Kutasov, `World Sheet and Space Time
Physics in Two Dimensional (Super) String Theory,' Princeton preprint
PUPT-1276 (September 1991)}
\REF\kutsei{D. Kutasov and N. Seiberg, \plb{251}{91}{67}; D. Kutasov,
G. Moore and N. Seiberg, unpublished}
\REF\mmsusy{E. Marinari and G. Parisi, \plb{240}{90}{375};
L. Alvarez-Gaum\'e and J.L. Ma\~nes, \mpl{6}{91}{2039};
A. Dabholkar, `Fermions and Nonperturbative Supersymmetry Breaking
in the One Dimensional Superstring,' Rutgers preprint RU-91/20 (May 1991);
A. Jevicki and J.P. Rodriques, `Supersymmetric Collective Field Theory,' Brown
preprint BROWN-HET-813 (June 1991)}
\REF\martinec{E. Martinec, unpublished}
\REF\wzwie{E. Witten and B. Zwiebach, `Algebraic Structures and Differential
Geometry in 2D String Theory,' IAS \&\ MIT preprint IASSNS-HEP-92/4 \&\
MIT-CTP-2057 (January 1992)}

%%%%%%%%%%%%%%%%%%%%%%%%%%%%%%%%%%%%%%%%%%%%%%%%%%%%%%%%%%%%%%%%%%%%%%%%%%%
%
\chapter{Introduction}

Since the discovery of the topological symmetry underlying string theory in
$c<1$ dimensions [\wtopgr --\dijk], it has become a common
feeling that the case of $c=1$ (see [\klebrev,\kutarev] for a review) might
share at least some topological properties with the simplest exactly solvable
models [\wsurp,\polyakov].  By naive counting, gravity in two dimensions has
minus one field-theoretical degrees of freedom, which suggests that at $c=1$
we are on the verge after which topological symmetry ceases to be an
explicit symmetry of the model.  Still, there is the long-standing conjecture
[\oldwitten] that quantum gravity and string theory may have an unbroken,
topological phase, even in dimensions higher than one.  However, for this
to be the case, topological symmetry should be hidden in the broken phase
in an as yet unknown manner, hence the issue of spontaneous breakdown of
topological symmetry would be crucial in this respect.

The theory of strings in two dimensions is presumably the simplest string
theory in which topological symmetry of any kind is not manifest.
The recent discovery [\wground] and analysis [\klebpol --\klebward] of the
ground ring structure in two dimensional string theory provides new support
to the idea advocated some time ago [\polyakov]
that discrete states [\discrstates] of $c=1$ string theory may have a
topological origin.  Discrete states have been analyzed thoroughly [\lizu]
in the framework of stringy BRST cohomologies.  Their correlation functions
have been discussed in [\danielss].  The discrete states are closely related
to spacetime $W_\infty$-like symmetries of the model
[\wground,\klebpol,\klebward,\danielss,\ajwinfty --\conewinf].

In this paper, we will discuss topological sigma model with the
two dimensional toroidal target.  The motivation for the present work
is the hope that the topological torus coupled to topological gravity might
represent a topological phase of two dimensional string theory.
This model has been touched upon from several
viewpoints in [\dvv,\wtoporb,\lgtorus];  some other topological aspects
of $c=1$ string theory have been studied in the context of the Penner model
in [\distvafa], and the Liouville approach in [\japan].

Possible connections with $c=1$ string theory do not represent the only
motivation for the study of the topological torus.  Indeed, the topological
torus is an excellent example of topological field theory leading to
interesting phenomena of their own.  It has an infinite number of physical
states, which lead to an enormous quantum symmetry of the model, and which
might be eventually assembled into a field theoretical degree of freedom.
Moreover, the topological torus represents one of the simplest topological
sigma models with $\pi_1(M)\neq 0$, a class of targets not yet fully
understood.  Another motivation comes from the fact that the model can be
easily extended to higher target dimensions, and may thus eventually shed
some light on the issue of higher symmetries in string theory [\highsym].

The paper is organized as follows.  In \S{2}, we set up the basic framework
of the topological torus sigma model on flat worldsheet, and identify
BRST invariant observables (\S{2.1}).  We analyze their structure at ghost
number zero in \S{2.2}, and at nonzero ghost numbers in \S{2.3}.
Point-like observables of ghost number zero form a topological ground ring,
which is extended by observables of nonzero ghost numbers to
a superring.  Bosonic generators of the ground ring span
a manifold, which we will refer to as the `spacetime' manifold, to distinguish
it from the `target' torus.  Fermionic generators extend the spacetime
manifold to a supermanifold.  Via descent equations, the ground ring
observables give rise to a symmetry algebra, formed by BRST invariant
conserved charges.  We study these symmetries in \S{3}.  In \S{3.1} we show
that the bosonic part of the symmetry algebra generates the algebra of all
spacetime diffeomorphisms (with coefficients in the topological ground ring),
while its supersymmetric extension preserves a symplectic form on the
spacetime supermanifold.  In \S{3.2} we study symmetries of the decompactified
limit in which the target torus becomes an infinite cylinder.  In this
limit, the symmetry algebra can be extended in a weak sense to the
topologically twisted $N=2$ superconformal algebra, hence realizing
topological BRST symmetry in spacetime.  The BRST charge of this spacetime
topological symmetry is conserved, but not BRST invariant under the
topological BRST charge on the worldsheet, hence the target topological
symmetry is hidden in the model.  \S{4} offers some summarizing discussion.

%%%%%%%%%%%%%%%%%%%%%%%%%%%%%%%%%%%%%%%%%%%%%%%%%%%%%%%%%%%%%%%%%%%%%%%%%%%%%

\chapter{The Topological Torus}

We will start with the following free field Lagrangian,
$$I_0 = \frac{1}{\pi}\intz (\p_z\bar X\p_{\bar z}X-\chi_z\p_{\bar z}
\psi -\bar\chi_{\bar z}\p_z\bar\psi ),
\eqn\ttlag$$
which describes the topological sigma model on flat worldsheet $\Sigma$ and
with a flat two dimensional torus as the target.  This is a conformal field
theory (CFT) even before imposing the BRST constraint, hence we will use
frequently some parts of CFT terminology.  In \ttlag , $X,\bar X$ are complex
coordinates on the torus, which has its K\"{a}hler structure fixed once and
for all.  Namely, we assume
$$X\equiv X+2\pi mR+2\pi nR\,\tau_0, \eqn\perix$$
with $R$ real, and $\im\tau_0\neq 0$.  $\psi ,\bar\psi$ of \ttlag\ are
fermionic ghosts of conformal weight zero, and the antighosts
$\chi_z ,\bar\chi_{\bar z}$ have conformal weights (0,1) and (1,0)
respectively.  On-shell we shall write
$$\eqalign{X(z,\bar z)=X(z)+X(\bar z),&\quad\bar X(z,\bar z)=\bar X(z)+
\bar X(\bar z),\cr
\psi (z),\quad\bar\psi (\bar z),&\quad\chi_z (z),\quad
\bar\chi_{\bar z}(\bar z).\cr}
\eqn\onshf$$
With this normalization, we have
$$\eqalign{X(z)\,\bar X(w)&\sim-\ln (z-w),\cr
\psi (z)\,\chi_w(w)&\sim -\frac{1}{z-w},\cr}\eqn\opef$$
and similarly for the right-movers.  The energy-momentum tensor of the
model is the standard one of free fields,
$$T_{zz}(z)=-\p_zX\p_z\bar X +\chi_z\p_z\psi, \eqn\enmomten$$
while the exotic form of the bosonic kinetic term in \ttlag\ only affects the
zero-mode part of $X$ and $\bar X$:
$$\eqalign{X(z,\bar z)&=x-i\left(\frac{in_1}{2R}+\frac{n_1\,\re\tau_0}{2R\,
\im\tau_0}+\frac{n_2}{2R\,\im\tau_0}+m_1R+m_2R\tau_0\right) \ln z \cr
&\qquad -i\left( \frac{in_1}{2R}+\frac{n_1\,\re\tau_0}{2R\,\im\tau_0}
+\frac{n_2}{2R\,\im\tau_0}\right) \ln \bar z + {\rm oscillators}.
\cr}\eqn\zeromx$$
Physical states will have $n_1,n_2=0$, hence only winding states will survive
the BRST condition.

%%%%%%%%%%%%%%%%%%%%%%%%%%%%%%%%%%%%%%%%%%%%%%%%%%%%%%%%%%%%%%%%%%%%%%%%%%%%
\section{Observables}

Topological BRST symmetry of \ttlag\ acts (on-shell) via its BRST charge
by
$$\eqalign{[Q, X]&=\psi,\cr \{ Q,\psi \}&=0,\cr
\{ Q,\chi_z \}&=\p_z\bar X,\cr }\qquad\qquad
\eqalign{[ Q,\bar X]&=\bar\psi,\cr  \{ Q,\bar\psi \}&=0,\cr
\{ Q,\bar\chi_{\bar z}\}&=\p_{\bar z} X.\cr} \eqn\ttbrst $$
As $X,\bar X$ do not exist as quantum fields, $\psi$ and $\bar\psi$ are
BRST nontrivial.  Hence, they give rise to the set of nontrivial point-like
observables that correspond to the homology ring of the target, quite
analogously as in any other topological sigma model.  We will refer to these
as the `homology observables,' and will denote them by
$$O^{(0)}_0=1, \qquad P^{(0)}_0=\psi, \qquad Q^{(0)}_0=\bar\psi, \qquad
R^{(0)}_0=\psi\bar\psi . \eqn\homolobs$$
On the other hand,
as the fundamental group of the target is nontrivial, we obtain a new set
of observables from winding sectors.%
\foot{This fact has been pointed out in [\dvv].}
{}From the worldsheet point of view, the existence of these new observables is
related to the fact that the following composites,
$$O^{(0)}_k(z,\bar z)=e^{ik\bar X(z)+i\bar kX(\bar z)} \eqn\homotobs$$
with $k=mR+nR\tau_0,\ \bar k=-mR-nR\bar\tau_0$,
carry conformal weight (0,0) with respect to the energy-momentum tensor
\enmomten\ (and its right-moving counterpart).  These observables can be
naturally referred to as `homotopy observables.'  General point-like physical
observables can be constructed as products of the homotopy
and homology observables.  We will denote the corresponding fields by
$$ P^{(0)}_k=\psi\; O^{(0)}_k, \qquad Q^{(0)}_k=\bar\psi\; O^{(0)}_k, \qquad
R^{(0)}_k=\psi\bar\psi\; O^{(0)}_k. \eqn\pointobs$$

In cohomological field theories [\wcohom], BRST invariant fields of conformal
weight (0,0) and ghost number $(m,n)$ give rise to a new set of observables,
with conformal weights $(p,q)$ and ghost number $(m-p,n-q)$.  The key
to this hierarchy of observables are the (on-shell) descent equations,
$$\eqalign{\{ Q, \CO^{(0)}\}&=0,\cr \{ Q, \CO^{(1)}\}&=d\CO^{(0)},\cr
\{ Q,\CO^{(2)}\}&=d\CO^{(1)},\cr}\eqn\descent $$
which can be split in the case at hand to the
left-moving and right-moving part.%
\foot{Actually, an even more refined splitting exists, as both of the exterior
derivatives involved (i.e.\ $d$ and $Q$) can be split into $d\equiv\p +\bar\p ,
\ Q= Q_L+Q_R$, and analogously for the ghost number.  We will mostly avoid
using this splitting in the following, however.}
(In \descent , $\CO^{(0)}$ is an arbitrary physical observable of
conformal weight (0,0), and $\{~\, ,~\}$ denotes either commutator or
anticommutator here.)

To summarize, we have encoutered a remarkably rich structure of observables
in the pure matter theory, even before coupling to gravity.  Thus,
we have worldsheet scalars
$$\eqalign{O^{(0)}_k(z,\bar z)&=e^{ik\bar X(z)+i\bar kX(\bar z)},\cr
P^{(0)}_k(z,\bar z)&=\psi\; e^{ik\bar X(z)i\bar kX(\bar z)},\cr
Q^{(0)}_k(z,\bar z)&=\bar\psi\; e^{ik\bar X(z)+i\bar kX(\bar z)},\cr
R^{(0)}_k(z,\bar z)&=\psi\bar\psi\; e^{ik\bar X(z)+i\bar kX(\bar z)}\cr}
\eqn\ghzero$$
of ghost numbers (0,0), (0,1), (1,0) and (1,1),
parametrized by $k$, the winding number on the target.
These fields have partners of conformal weight (1,0) with ghost numbers
(--1,0), (0,0), (--1,1) and (0,1), and partners of conformal weight (0,1) with
ghost numbers (0,--1), (0,0), (1,--1) and (1,0), assembled in the following
one-forms:
$$\eqalign{O^{(1)}_k(z,\bar z)&=i(k\chi +\bar k\bar\chi )\; e^{ik\bar X(z)+
i\bar kX(\bar z)},\cr
P^{(1)}_k(z,\bar z)&=i(i\p X+k\psi\chi +\bar k\psi\bar\chi )\;
e^{ik\bar X(z)+i\bar kX(\bar z)},\cr
Q^{(1)}_k(z,\bar z)&=i(i\bar\p\bar X+\bar k\bar\psi\bar\chi +k\bar\psi\chi )\;
e^{ik\bar X(z)+i\bar kX(\bar z)},\cr
R^{(1)}_k(z,\bar z)&=i\left( (i\p X+k\psi\chi )\bar\psi
+\psi(i\bar\p\bar X+\bar k\bar\psi\bar\chi )\right)
e^{ik\bar X(z)+i\bar kX(\bar z)}.\cr}
\eqn\ghone$$
(Here and throughout, we use the following notation: $\p =dz\,\p_z,\
\bar\p =d\bar z\,\p_{\bar z},\ \chi =\chi_z\,dz,\ \bar\chi =\bar\chi_{\bar z}\,
d\bar z.$)
Moreover, partners of \ghzero\ exist with conformal weight (1,1) and
ghost numbers (--1,--1), (--1,0), (0,--1) and (0,0),
$$\eqalign{O^{(2)}_k(z,\bar z)&=-k\bar k\;\chi\wedge\bar\chi\; e^{ik\bar X(z)+
i\bar kX(\bar z)},\cr
P^{(2)}_k(z,\bar z)&=-(i\p X+k\psi\chi )\wedge\bar k\bar\chi\;
e^{ik\bar X(z)+i\bar kX(\bar z)},\cr
Q^{(2)}_k(z,\bar z)&=-(i\bar\p\bar X+\bar k\bar\psi\bar\chi )\wedge k\chi\;
 e^{ik\bar X(z)+i\bar kX(\bar z)},\cr
R^{(2)}_k(z,\bar z)&=-(i\p X+k\psi\chi )\wedge (i\bar\p\bar X+
\bar k\bar\psi\bar\chi )\; e^{ik\bar X(z)+i\bar kX(\bar z)}.\cr}
\eqn\ghtwo$$
Each of these fields will play its specific role below.

%%%%%%%%%%%%%%%%%%%%%%%%%%%%%%%%%%%%%%%%%%%%%%%%%%%%%%%%%%%%%%%%%%%%%%%%%%%%%%
\section{Ghost Number Zero}

Now we will show that an important structure exists already at ghost number
zero.  Indeed, observables $O^{(0)}_k$ with ghost number zero and conformal
weight (0,0) form a ring which, as will be apparent below, is the topological
analog of the ground ring of two dimensional string theory discovered in
[\wground].%
\foot{More precisely, it is the matter part of the corresponding analog, as
the Liouville dimension has not yet been included.  The full-fledged analog
should come from the coupling to topological gravity.}
We have one
element of this ring for each element of the lattice corresponding to the
toroidal target.  Hence, the ring is free and commutative, with two
generators
$$\eqalign{a&=e^{ik_a\bar X(z)+i\bar k_aX(\bar z)},\cr
b&=e^{ik_b\bar X(z)+i\bar k_bX(\bar z)},\cr}\eqn\genring$$
where $k_a,k_b$ is an arbitrarily chosen basis of the torus' lattice.  Generic
elements of the ring are finite sums of
$$a^mb^n,\eqn\genelem$$
with $m,n\in {\bf Z}$.  In geometrical terms, the ring is isomorphic to
the group ring of the fundamental group of the target manifold.  We will
refer to it as the `topological ground ring' of the sigma model, and denote
it by $\CR$.

$P^{(1)}_k$ and $Q^{(1)}_k$ are one-forms on the worldsheet.
Each of them is conserved modulo BRST commutators by virtue of
the descent equation \descent .  As the one-forms carry ghost number zero,
they generate an algebra of  symmetries of the topological ground ring.
This algebra can be identified from the operator product expansion (OPE)
of these currents (with $dz,d\bar z$ omitted):
$$P^{(1)}_k(z,\bar z)\, P^{(1)}_\ell (w,\bar w)\sim
i(\ell -k)\frac{P^{(1)}_{k+\ell}(w,\bar w)}{z-w},\eqn\algope$$
and similarly for $Q^{(1)}_k$'s, with $(\bar\ell -\bar k)$ replacing
$(\ell - k)$ on the right hand side.

The corresponding charges
$$L_k\equiv\frac{1}{2\pi}\oint_CP^{(1)}_k(z,\bar z)\eqn\vircharg$$
form an infinite algebra, which is surprisingly close to the Virasoro algebra:
$$[L_k,L_\ell ]=(k-\ell )L_{k+\ell}. \eqn\vira$$
The action of the algebra on the topological ground ring is encoded
in the following OPE:
$$P^{(1)}_k(z,\bar z)\, O^{(0)}_\ell (w,\bar w)\sim
i\ell\;\frac{O^{(0)}_{k+\ell}(w, \bar w)}{(z-w)}.\eqn\actope$$
As the topological ground ring is the ring of (positive as well as negative)
power series in two generators \genring , it is sufficient to identify the
action of the Virasoro-like algebra \vira\ on the generators $a,b$ of $\CR$.
{}From \actope\ we obtain
$$\eqalign{L_{mR+nR\tau_0}\cdot a&=-R\; a^{m+1}b^n,\cr
L_{mR+nR\tau_0}\cdot b&=-R\tau_0\; a^mb^{n+1}.\cr}\eqn\actgen$$
Hence, the Virasoro-like algebra \vira\ acts on the $(a,b)$ plane via specific
infinitesimal diffeomorphisms, with
$$L_{mR+nR\tau_0}=-R\; a^{m+1}b^n\frac{\p}{\p a}-R\tau_0\; a^mb^{n+1}
\frac{\p}{\p b}, \eqn\viraact$$
while the analogous algebra generated by $\bar L_k$,
$$\bar L_k\equiv\frac{1}{2\pi}\oint_CQ^{(1)}_k(z,\bar z), \eqn\cvircharg$$
acts by
$$\bar L_{mR+nR\tau_0}=-R\; a^{m+1}b^n\frac{\p}{\p a}-R\bar\tau_0\; a^mb^{n+1}
\frac{\p}{\p b}\eqn\cviraact$$
and satisfies
$$[\bar L_k,\bar L_\ell ]=-(\bar k-\bar\ell )\bar L_{k+\ell}.\eqn\cvira$$
Commutation relations between $L_k$ and $\bar L_\ell$,
$$[L_k,\bar L_\ell ]=-\bar k\,L_{k+\ell}-\ell\,\bar L_{k+\ell},\eqn\cvira$$
can be determined either from the corresponding OPE,
$$P^{(1)}_k(z,\bar z)\, Q^{(1)}_\ell (w,\bar w)\sim
i\ell\,\frac{Q^{(1)}_{k+\ell}(w,\bar w)}{z-w}-i\bar k\,
\frac{P^{(1)}_{k+\ell}(w,\bar w)}{\bar z-\bar w},\eqn\opecvira$$
or by commuting directly \viraact\ and \cviraact\ on the $(a,b)$ plane.

Finally, $R^{(2)}_k$, which are two-forms of ghost number zero, can be
multiplied by their own coupling constants and added to the Lagrangian
\ttlag ,
$$I (\alpha_k)=I_0+\sum_k\alpha_k\int_\Sigma R^{(2)}_k(z,\bar z),
\eqn\deflzero$$
hence deforming the theory while preserving topological symmetry.  We
must be careful here however, as it is still necessary to check which
values of $\alpha_k$ are permitted, i.e.\ which $R^{(2)}_k$'s are genuine
moduli of the theory.  In particular, it is important to bear in mind that
under these deformations, the on-shell structure we have studied so far may
be changed.

The possibility of adding terms $R^{(2)}_k$ with nonzero $k$ to the
Lagrangian is extremely attractive, as they break translational invariance
in the target.  On the other hand, this possiblility is presumably not
related to the non-conservation of momenta in Liouville theory, as the
topological analog of Liouville dimension is expected to emerge just after
coupling to topological gravity.  Note that even the deformed Lagrangian
\deflzero\ still carries ghost number (0,0).

%%%%%%%%%%%%%%%%%%%%%%%%%%%%%%%%%%%%%%%%%%%%%%%%%%%%%%%%%%%%%%%%%%%%%%%%%%%%%
\section{Nonzero Ghost Numbers}

At ghost number one, we have two infinite sequences of fermionic
point-like observables, namely $P^{(0)}_k$ and $Q^{(0)}_k$ of \pointobs ,
each of them parametrized by target winding numbers.  Together with the
bosonic observables $R^{(0)}_k$ of ghost number (1,1), they comprise an
extension of the topological ground ring $\CR$ to a supersymmetric version
thereof, which we will denote by $\CR '$.  The following OPEs between the
observables of conformal weight (0,0),
$$\eqalign{O^{(0)}_k(z,\bar z)\, P^{(0)}_\ell (w,\bar w)&\sim P^{(0)}_{k+\ell}
(w,\bar w),\cr
O^{(0)}_k(z,\bar z)\, Q^{(0)}_\ell (w,\bar w)&\sim Q^{(0)}_{k+\ell}
(w,\bar w),\cr
O^{(0)}_k(z,\bar z)\, R^{(0)}_\ell (w,\bar w)&\sim R^{(0)}_{k+\ell}
(w,\bar w),\cr
P^{(0)}_k(z,\bar z)\, Q^{(0)}_\ell (w,\bar w)&\sim R^{(0)}_{k+\ell}
(w,\bar w),\cr}
\eqn\ssring$$
and zero otherwise, show that this fermionic extension of $\CR$ is
a free ring generated multiplicatively by the generators $a,b$ of $\CR$
(with both positive and negative powers permitted),
together with two anticommuting generators $\theta,\bar\theta$,
$$\theta\equiv\psi (z),\qquad\bar\theta\equiv\bar\psi (\bar z).\eqn\defthe$$
Geometrically, this ring is naturally
isomorphic to the tensor product of the homology ring of the target
and the group ring of its fundamental group.  Elements of $\CR '$
are finite sums of
$$a^m\, b^n\,\theta^p\,\bar\theta^q, \qquad m,n\in{\bf Z},\quad p,q\in\{0,1\}.
\eqn\ringgen$$

Analogously, the algebra of symmetries of the topological ground ring $\CR$
is extended to a superalgebra of symmetries of the extended ring $\CR '$.
We will define the charges that correspond to nonzero ghost number currents
of \ghone\ by
$$\eqalign{\CQ_k&=\frac{1}{2\pi}\oint_CO^{(1)}_k(z,\bar z),\cr
\CG_k&=\frac{1}{2\pi}\oint_CR^{(1)}_k(z,\bar z),\cr}\eqn\ghcharges$$
and determine their algebra from the OPEs of the currents.
\foot{In the remainder of this section, we will work in the decompactification
limit of ${\rm Im}\;\tau_0\rightarrow\infty$, which leads to $k=mR,\ \bar k
=-mR$.  We can set $k_a=R$, and $b$ is effectively zero.
The objects will be scaled such that $R=1$.}
First of all, tensorial properties of the new charges with respect to the
Virasoro-like algebra \vira\ and \cvira\ can be determined from
$$\eqalign{P^{(1)}_k(z,\bar z)\, O^{(1)}_\ell (w,\bar w)&\sim i\ell\;
\frac{O^{(1)}_{k+\ell}(w,\bar w)}{z-w},\cr
P^{(1)}_k(z,\bar z)\, R^{(1)}_\ell (w,\bar w)&\sim i(\ell -k)\,
\frac{R^{(1)}_{k+\ell}(w,\bar w)}{z-w}\cr}\eqn\anttens$$
(and analogously for $P^{(1)}_k$ replaced by $Q^{(1)}_k$, with some obvious
conjugations), which shows that $\CQ_m$ behave like Fourier components of
a one-form, while $\CG_m$ comprise a two-tensor with respect to \vira\ and
\cvira .  The anticommutation relation of the fermionic charges is encoded in
$$O^{(1)}_k(z,\bar z)\, R^{(1)}_\ell (w,\bar w)\sim
ik\;\frac{Q^{(1)}_{k+\ell}(w,\bar w)}{z-w}-i\bar k\;\frac{P^{(1)}_{k
+\ell}(w,\bar w)}{\bar z-\bar w}.\eqn\antiope$$
We can thus obtain the complete algebra of quantum symmetries of the
topological ground ring $\CR'$, as generated by BRST invariant currents
\ghone , in the decompactification limit studied here.  After redefining
$$\eqalign{\CL_m&=\half (L_m+\bar L_m),\cr
\CJ_m&=\bar L_m-L_m,\cr}\eqn\bosgen$$
we arrive at the following algebra of charges:
$$\eqalign{[\CL_m,\CL_n]&=(m-n)\CL_{m+n},\cr [\CL_m,\CQ_n]&=-n\CQ_{m+n},\cr
[\CL_m,\CG_n]&=(m-n)\CG_{m+n},\cr \{\CQ_m,\CG_n\}&=-m\CJ_{m+n},\cr}
\qquad\qquad
\eqalign{[\CJ_m,\CJ_n]&=0,\cr [\CJ_m,\CQ_n]&=0,\cr [\CJ_m,\CG_n]&=0,\cr
[\CL_m,\CJ_n]&=-n\CJ_{m+n}.\cr}\eqn\qalgcr$$
This is an extension of the Virasoro-like algebra \vira\ by fermionic
generators.  We will analyze its possible physical meaning in the next
section.

By definition, the algebra \qalgcr\ acts on the fermionic extension of the
ground ring, $\CR '$, as a symmetry algebra.  As $\CR '$ is free and
finitely generated, the action of the symmetry algebra is determined by
its action on the superplane of $(a,\theta , \bar\theta)$, where it acts
by (infinitesimal) super-diffeomorphisms.  From the corresponding OPEs, of
which the first few examples are
$$\eqalign{P^{(1)}_k(z,\bar z)P^{(0)}_\ell(w,\bar w)&\sim i(\ell -k)
\frac{P^{(0)}_{k+\ell}(z,\bar z)}{z-w},\cr
P^{(1)}_k(z,\bar z)\, Q^{(0)}_\ell(w,\bar w)&\sim i\ell\;\frac{Q^{(0)}_{k+\ell}
}{z-w} -i\bar k\;\frac{P^{(0)}_{k+\ell}(z,\bar z)}{\bar z-\bar w},\cr}
\eqn\supact$$
we infer
$$\eqalign{\CL_m&=-a^{m+1}\frac{\p}{\p a}+\half m\, a^m(\theta +\bar\theta )
\left( \frac{\p}{\p\theta}+\frac{\p}{\p\bar\theta}\right),\cr
\CJ_m&=m\, a^m(\bar\theta -\theta )\left( \frac{\p}{\p\theta}+\frac{\p}{\p
\bar\theta}\right) ,\cr
\CQ_m&=m\, a^m\left( \frac{\p}{\p\theta}+\frac{\p}{\p\bar\theta}\right),\cr
\CG_m&=-a^{m+1}(\bar\theta -\theta )\frac{\p}{\p a}-m\,\theta\bar\theta\, a^m
\left( \frac{\p}{\p\theta}+\frac{\p}{\p\bar\theta}\right) .\cr}\eqn\supvecact
$$
The formulae obtained here can be easily extended to the full-fledged torus
with both of the target dimensions compactified, as we will see below.
We will discuss the decompactification limit again in \S{3.2}.

The analysis of the observables with nonzero ghost numbers is completed by
mentioning the existence of two-forms woth nonzero ghost numbers in \ghtwo .
At ghost number (--1,--1) we have two-forms $O^{(2)}_k$, which are
BRST invariant up to total derivative.  Hence their integrals over $\Sigma$
are physical observables, and can be used to deform the Lagrangian to
$$I (\alpha_k,\beta_\ell )=I_0+\sum_k\alpha_k\int_\Sigma R^{(2)}_k(z,
\bar z)+\sum_\ell\beta_\ell\int_\Sigma O^{(2)}_\ell (z,\bar z).\eqn\deflag$$
The ghost number is no longer conserved if at least some of the $\beta_\ell$'s
are nonzero.  $P^{(2)}_k$ and $Q^{(2)}_k$ are two-forms as well, but cannot
be simply added to the Lagrangian, as they are fermionic.
%

%%%%%%%%%%%%%%%%%%%%%%%%%%%%%%%%%%%%%%%%%%%%%%%%%%%%%%%%%%%%%%%%%%%%%%%%%%%%%
%
\chapter{Spacetime Symmetries}

%%%%%%%%%%%%%%%%%%%%%%%%%%%%%%%%%%%%%%%%%%%%%%%%%%%%%%%%%%%%%%%%%%%%%%%%%%%%%
\section{Spacetime Diffeomorphisms}

In the previous  subsection, we have mainly analyzed the theory in the
decompactified limit of $\im\tau_0\rightarrow\infty$.  Now we will study
the full-fledged model.

The full algebra of symmetries is now
$$\eqalign{[L_k,L_\ell ]&=(k-\ell )L_{k+\ell},\cr [L_k,\CQ_\ell ]&=
-\ell\,\CQ_{k+\ell},\cr
[L_k,\CG_\ell ]&=(k-\ell )\CG_{k+\ell},\cr \{\CQ_k,\CG_\ell\}&=
-k\,\bar L_{k+\ell}-\bar k\, L_{k+\ell},\cr}
\qquad\qquad
\eqalign{[\bar L_k,\bar L_\ell ]&=-(\bar k-\bar\ell )\bar L_{k+\ell},\cr
[\bar L_k,\CQ_\ell ]&=\bar\ell\,\CQ_{k+\ell},\cr [\bar L_k,\CG_\ell ]&=
-(\bar k-\bar\ell )\CG_{k+\ell},\cr
[L_k,\bar L_\ell ]&=-\bar k\, L_{k+\ell}-\ell\,\bar L_{k+\ell}.\cr}
\eqn\qalgff$$
This algebra acts on the superplane of generators of the topological ground
ring $(a,b,\theta ,\bar\theta )$ by the following vector fields,
$$\eqalign{L_k&=-R\, a^{m+1}b^n\frac{\p}{\p a}-R\tau_0\, a^mb^{n+1}
\frac{\p}{\p b}+ R\, a^mb^n\theta \left( (m+n\tau_0)\frac{\p}{\p\theta}
+(m+n\bar\tau_0)\frac{\p}{\p\bar\theta}\right) ,\cr
\bar L_k&=-R\, a^{m+1}b^n\frac{\p}{\p a}-R\bar\tau_0\, a^mb^{n+1}
\frac{\p}{\p b}+ R\, a^mb^n\bar\theta \left( (m+n\tau_0)\frac{\p}{\p\theta}
+(m+n\bar\tau_0)\frac{\p}{\p\bar\theta}
\right) ,\cr
\CG_k&=R(\theta -\bar\theta )a^{m+1}b^n\frac{\p}{\p a}+R(\bar\tau_0\theta
-\tau_0\bar\theta )a^mb^{n+1}\frac{\p}{\p b}\cr
&\qquad \qquad \qquad \qquad \qquad \qquad \qquad \ \
-R\, a^mb^n\theta\bar\theta\left( (m+n\tau_0)\frac{\p}{\p\theta}
+(m+n\bar\tau_0)\frac{\p}{\p\bar\theta}\right) ,\cr
\CQ_k&=R\, a^mb^n\left( (m+n\tau_0)\frac{\p}{\p\theta}
+(m+n\bar\tau_0)\frac{\p}{\p\bar\theta}\right) ,\cr}\eqn\supvecff$$
where we have set $k\equiv mR+nR\tau_0,\ k_a\equiv R,\ k_b\equiv R\tau_0$.

We can get some better insight into the structure of the symmetry algebra
after switching to its basis with real structure constants, first by
performing the fermionic coordinate change to
$$\Theta_a=\frac{\tau_0\bar\theta -\bar\tau_0\theta}{\tau_0 -\bar\tau_0},
\qquad \Theta_b=\frac{\theta -\bar\theta}{\tau_0 -\bar\tau_0},\eqn\newfermi$$
where $\Theta_a,\Theta_b$ is the natural basis of the first integral
cohomology group of the target, and then by switching from $L_k,\ldots ,\CG_k$
to
$$\eqalign{\hat\CL^a_{m,n}&=\frac{\tau_0\bar L_{mR+nR\tau_0}
-\bar\tau_0L_{mR+nR\tau_0}}{R(\tau_0 -\bar\tau_0)},\cr
\hat\CG_{m,n} &=\frac{\CG_{mR+nR\tau_0}}{R(\tau_0 -\bar\tau_0)},\cr}
\qquad
\eqalign{\hat\CL^b_{m,n}&=\frac{L_{mR+nR\tau_0}-\bar L_{mR+nR
\tau_0}}{R(\tau_0 -\bar\tau_0)},\cr
\hat\CQ_{m,n}&=\frac{\CQ_{mR+nR\tau_0}}{R}.\cr}\eqn\redefchar$$
After these changes, the vector field representation \supvecff\ simplifies
to
$$\eqalign{\hat\CL^a_{m,n}&=-a^{m+1}b^n\frac{\p}{\p a}+a^mb^n\Theta_a
\left( m\frac{\p}{\p\Theta_a}+n\frac{\p}{\p\Theta_b}\right),\cr
\hat\CL^b_{m,n}&=-a^mb^{n+1}\frac{\p}{\p b}+a^mb^n\Theta_b \left(
m\frac{\p}{\p\Theta_a}+n\frac{\p}{\p\Theta_b}\right),\cr
\hat\CG_{m,n}&=a^{m+1}b^n\Theta_b\frac{\p}{\p a}-a^mb^{n+1}\Theta_a\frac{\p}{
\p b}+a^mb^n\Theta_a\Theta_b\left( m\frac{\p}{\p\Theta_a}+
n\frac{\p}{\p\Theta_b}\right),\cr
\hat\CQ_{m,n}&=a^mb^n\left( m\frac{\p}{\p\Theta_a}+n\frac{\p}{\p\Theta_b}
\right) .\cr}\eqn\redefvecff$$
In terms of the redefined generators \redefchar , the bosonic part of the
symmetry algebra \qalgff\ takes the following form:
$$\eqalign{[\hat\CL^a_{m,n},\hat\CL^a_{p,q}]&=(m-p)\hat\CL^a_{m+p,n+q},\cr
[\hat\CL^b_{m,n},\hat\CL^b_{p,q}]&=(n-q)\hat\CL^b_{m+p,n+q},\cr
[\hat\CL^a_{m,n},\hat\CL^b_{p,q}]&=n\,\hat\CL^a_{m+p,n+q}-
p\,\hat\CL^b_{m+p,n+q}.\cr}\eqn\redefqalgffb$$
When restricted to the action on $(a,b)$, \redefqalgffb\ is the full algebra
of infinitesimal (ground ring valued) diffeomorphisms of the $(a,b)$ plane.
However, each generator of the algebra \redefvecff\ has an important
fermionic part, proportional to $m\,\p /\p\Theta_a+n\,\p /\p\Theta_b$.  In
the superplane of $(a,b,\Theta_a,\Theta_b)$, the algebra \redefqalgffb\ and
its supersymmetric extension to
$$\eqalign{
\eqalign{[\hat\CL^a_{m,n},\hat\CG_{p,q}]&=(m-p)\hat\CG_{m+p,n+q},\cr
[\hat\CL^a_{m,n},\hat\CQ_{p,q}]&=-p\,\hat\CQ_{m+p,n+q},\cr}&\qquad\qquad
\eqalign{[\hat\CL^b_{m,n},\hat\CG_{p,q}]&=-(n-q)\hat\CG_{m+p,n+q},\cr
[\hat\CL^b_{m,n},\hat\CQ_{p,q}]&=-q\,\hat\CQ_{m+p,n+q},\cr}\cr
\{\hat\CQ_{m,n},\hat\CG_{p,q}\}&=-n\,\hat\CL^a_{m+p,n+q}+m\,
\hat\CL^b_{m+p,n+q}\cr}
\eqn\redefqalgfff$$
acts by some specific superdiffeomorphisms.  In this way, the algebra of all
diffeomorphisms of the $(a,b)$ plane gets extended to an algebra of volume
preserving superdiffeomorphisms, where the preserved volume on the superplane
is given by
$${\rm Vol}=\frac{da}{a}\wedge\frac{db}{b}\wedge d\Theta_a\wedge d\Theta_b.
\eqn\supvol$$
Moreover, it is easy to observe that \redefvecff\ does not exhaust the whole
set of all Vol-preserving superdiffeomorphisms.  Actually,
our symmetry algebra preserves in addition to \supvol\ also the following
symplectic two-form
$$\omega=\frac{da}{a}\wedge d\Theta_a+\frac{db}{b}\wedge d\Theta_b
\eqn\symplform$$
on the superplane.  This suggests that a relation to grassmannian particle
mechanics might exist:  If we interpret the superplane $(a,b,\Theta_a,
\Theta_b)$ as the phase space of a mechanical system with two degrees of
freedom evolving in a purely grassmannian time coordinate, then
\symplform\ can be considered the canonical symplectic form of the system.

In the conventional phase of string theory in two dimensions, classical limit
of string theory corresponds to the one-particle limit of the free fermions
in the target [\polch], and Witten's ground ring is an algebra of polynomial
functions on the extended phase space of the particle.  The corresponding
symmetry algebra preserves the symplectic structure on the phase space.
When compared with the observations just made, we see that in the matter
sector of the topological string theory on the torus, we have arrived at
another one-particle system: the grassmannian particle in two dimensions.
This correspondence bears some remote resemblance to $c=-2$ matrix models
[\minustwo], where the starting point is essentially a grassmannian matrix
mechanics.

Note that the algebra of vector fields \redefvecff\ enjoys an important
symmetry.  After denoting
$$D_a\equiv \frac{\p}{\p\Theta_a},\qquad\qquad D_b\equiv \frac{\p}{\p\Theta_b},
\eqn\fermider$$
we observe that the full algebra can be generated from $\hat\CG$ by
consecutive (anti)commutations with $D$'s.  Schematically,
$$\eqalign{\{ D_a,\hat\CG_{m,n}\}&=\hat\CL^b_{m,n},\cr
[ D_b,\hat\CL^b_{m,n}]&=\hat\CQ_{m,n} ,\cr}\qquad
\eqalign{\{ D_b,\hat\CG_{m,n}\}&=-\hat\CL^a_{m,n},\cr
[D_a,\hat\CL^a_{m,n}]&=\hat\CQ_{m,n}.\cr}\eqn\suppropalg$$
This leads to a succinct superspace formulation of the algebra, which we
won't enter into here.

Instead, note that our symmetry algebra \redefvecff\ does not generate
the full algebra of (ground ring valued) $\omega$-preserving
superdiffeomorphisms.  Actually, \redefvecff\ is the algebra of all
{\it exact} $\omega$-preserving diffeomorphisms, i.e.\ those diffeomorphisms
that are generated from Hamiltonians.  To obtain the full algebra of all
symplectomorphisms, two new generators, $D_a,D_b$ in the notation of
\fermider , should be added.  These two generators can be thought of as
the missing zero modes of $\hat\CQ$, as we have formally
$$D_a=\lim_{m\rightarrow 0}\frac{1}{m}\hat\CQ_{m,0},\qquad
D_b=\lim_{m\rightarrow 0}\frac{1}{m}\hat\CQ_{0,m}.\eqn\limzero$$
In terms of the underlying conformal field theory, these operators actually
do exist, and are given by
$$D_a=\frac{i}{2\pi}\oint_C(\chi (z)-\bar\chi (\bar z)) \qquad
D_b=\frac{i}{2\pi}\oint_C(\tau_0\chi (z)-\bar\tau_0\bar\chi (\bar z)).
\eqn\cftds$$
Clearly, they are not BRST invariant.  It is interesting, however, to note
that they are BRST invariant on the instanton moduli space of the
model, where $\p \bar X=0$.

%%%%%%%%%%%%%%%%%%%%%%%%%%%%%%%%%%%%%%%%%%%%%%%%%%%%%%%%%%%%%%%%%%%%%%%%%%%%%
\section{Spacetime Topological Symmetry}

Now we will return to the analysis of the symmetry algebra in the
decompactification limit of $\im\tau_0\rightarrow\infty$.  This limit not only
simplifies some expressions, but reveals an interesting (albeit
speculative) physical structure.

{}From the spacetime point of view, the symmetry algebra \qalgcr\ represents
a fermionic extension of the Virasoro algebra.  It is convenient to
introduce a new formal variable $Z$, and define
$$\eqalign{\CT (Z)&=\sum_m\CL_mZ^{-m-2},\cr
\CQ (Z)&=\sum_m\CQ_mZ^{-m-1},\cr}
\qquad
\eqalign{\CJ (Z)&=\sum_m\CJ_mZ^{-m-1},\cr
\CG (Z)&=\sum_m\CG_mZ^{-m-2}.\cr}
\eqn\tarcurr$$
The commutation relations \qalgcr\ can now be formally rewritten as OPEs in
this new variable.
Conformal weights of $\CQ (Z)$ and $\CG (Z)$ with respect
to $\CT (Z)$ are one and two respectively, which is the situation
encountered in the topologically twisted $N=2$ superconformal algebra in
two dimensions.  We are thus tempted to identify the symmetry algebra
of the decompactified model, \qalgcr , as the algebra of
spacetime topological symmetry.  There are important differences between
\qalgcr\ and the two dimensional topological algebra, however.
First of all, the zero mode of the natural candidate for the BRST current,
$\CQ (Z)$, does not exist.  As a result of this, $\CQ_n$ and  $\CG_m$
cannot anticommute to $\CL_{m+n}$, and a peculiar modification of the
topological symmetry algebra in spacetime would result.

To obtain a closer correspondence to the standard topological symmetry
algebra, we will have to go beyond the algebra of exact $\omega$-preserving
diffeomorphisms, which is generated by the charges of \S{2}, and invoke
one of the non-exact symplectomorphisms of \fermider .  Indeed, the
(rescaled) operator $D_b$, which we will now denote by
$$\BQ\equiv \frac{\p}{\p\bar\theta}-\frac{\p}{\p\theta},\eqn\toptarch$$
completes the algebra of $\CL_m$ and $\CG_n$ to the topological symmetry
algebra in spacetime,
$$\eqalign{[\CL_m,\CL_n]&=(m-n)\CL_{m+n},\cr
[\CL_m,\BQ ]&=0,\cr}\qquad\qquad
\eqalign{[\CL_m,\CG_n]&=(m-n)\CG_{m+n},\cr
\{\BQ ,\CG_n\}&=2\CL_{n}, \cr}\eqn\topalgweak$$
and plays the role of the BRST-like charge of this spacetime topological
symmetry.  In the underlying worldsheet CFT we have
$$\BQ =\frac{1}{2\pi}\oint\left( \chi (z)+\bar\chi (\bar z)\right) .
\eqn\cfttoptar$$
The worldsheet current of $\BQ$ does not enter the descent equations of $Q$
studied in \S{2}, as its $Q$-commutator is not equal to $d({\rm something})$.
Consequently, its charge, albeit conserved (on shell), is not physical with
respect to the worldsheet topological BRST symmetry.
(One more refinement, mentioned in \S{2.1}, should be checked here.  In
topological CFT, one can make use of the chiral splitting of the topological
BRST charge, $Q=Q_L+Q_R$, to define another nilpotent charge,
$$\tilde{Q}\equiv Q_L-Q_R.\eqn\chirbrst$$
In the case of the topological torus,
the topological ground ring is the ring of cohomology classes of both $Q$
and $\tilde{Q}$.  The new nilpotent charge defines its own descent
equations, hence its own class of currents conserved modulo $\tilde{Q}$ and
acting as symmetries on $\CR'$.  However, $\BQ$ does not enter the
descent equations of $\tilde{Q}$ either.)

The fact that the spacetime topological BRST charge is not invariant under
the worldsheet topological BRST, yet can it be defined as a conserved
charge, may suggest that the spacetime topological symmetry is still present
in the model, but is perhaps broken spontaneously.

With the interpretation of $\BQ$ as the BRST-like operator of
spacetime topological symmetry, it is natural to ask what do the physical
states of this cohomology operator, satisfying
$$\BQ |{\rm phys}\rangle =0.\eqn\phystartop$$
correspond to.  Despite the physical answer to this question is not completely
clear, the condition of topological BRST invariance in spacetime acquires a
natural geometrical meaning:
One linear combination of the topological ghosts $\psi$ and $\bar\psi$
corresponds to the would-be cohomology class in the decompactified
dimension of the target, and physical states in the sense of \phystartop\
are precisely those states that are independent of this linear combination
of ghosts -- indeed a very plausible condition in the decompactified limit.

In this subsection, we have tried to obtain at least some preliminary
interpretation of \qalgcr\ in terms of a topological
symmetry in spacetime.  Alternatively, we might try to interpret \qalgcr\
as a supersymmetry algebra in its more conventional sense.
Indeed, \qalgcr\ resembles the symmetry of the non-critical two dimensional
superstring, as studied in [\kutarev,\cernground,\difrakut,\kutsei]
(see also [\mmsusy]).  Actually, two dimensional
superstring theory with the chiral GSO projection has been conjectured
[\difrakut,\martinec] to be a topological theory.  As a result of two
dimensional kinematics, the spacetime supersymmetry charge $Q_{\rm susy}$
is nilpotent, which suggests that it can be used for imposing a BRST-like
condition on physical states,
$$Q_{\rm susy}|{\rm phys}\rangle =0,\eqn\susytopo$$
as has been pointed out in [\difrakut].
In such a model, the tachyon is essentially projected out by the GSO
projection and the BRST condition, hence only the  discrete (special) states
survive.  The analogy between the topological torus and superstring theory
in two dimensions deserves further investigation.

%%%%%%%%%%%%%%%%%%%%%%%%%%%%%%%%%%%%%%%%%%%%%%%%%%%%%%%%%%%%%%%%%%%%%%%%%%%%%

\chapter{Discussion}

In this paper we have analyzed symmetries of topological string theory on the
two dimensional torus.  We have found an interesting structure even before
the topological torus is coupled to topological gravity.

Point-like observables in the sigma model form a closed ring under operator
multiplication, which is conjectured to be the topological analog of the
ground ring of two dimensional string theory.

The model contains an infinite number of conserved BRST invariant
quantum currents.  The corresponding charges form a supersymmetric algebra of
symmetries of the topological ground ring.  A bosonic subsector of this large
symmetry algebra generates all (ground ring valued) diffeomorphisms of a
two dimensional spacetime.  Hence, the invariance of the model under the
symmetry algebra may signalize that, after coupling to gravity,  the model
does realize an unbroken phase of general relativity in spacetime.  Unlike
in generic topological sigma models, where the invariance under target
diffeomorphisms is usually proved as an independence of the target metric
with the use of the topological BRST symmetry on the worldsheet, here we
have obtained a direct realization of spacetime diffeomorphisms in terms
of conserved charges corresponding to quantum currents on the worldsheet.

The appearance of spacetime diffeomorphisms in the set of conserved charges
of the model is by no means obvious, and cannot be extended to arbitrary
topological sigma models.  This enormous quantum symmetry is related closely
to the nontrivial fundamental group of the target.  Indeed, the first
homology group of the target, which is simply
$$\pi_1(M)/[\pi_1(M),\pi_1(M)]\otimes {\bf R},$$
gives rise to the bosonic, ghost-number-preserving part of the symmetry
algebra of the model.  Second, $\pi_1(M)$ itself generates its own class of
observables, called homotopy observables in \S{2}, which make the dimension
of the space of physical states as well as the dimension of the symmetry
algebra infinite.

This realization of spacetime diffeomorphism invariance in terms of conserved
charges on the worldsheet might be relevant to string theory also on
more general grounds.  Note that the restriction to two target dimensions
made in this paper is not necessary, the model can be easily extended to
an arbitrary (even) number of target dimensions.  In this version of
topological string theory with a toroidal target, we will obtain, after
repeating the steps of \S{2} and \S{3}, the algebra of spacetime
diffeomorphisms as the algebra of BRST invariant conserved worldsheet
charges of ghost number zero.

At nonzero ghost numbers, the symmetry algebra extends to the algebra of
(ground ring valued) diffeomorphisms that preserve a symplectic form on
the superextension of the spacetime manifold, and are exact in the sense
that they are generated by Hamiltonians.  This algebra can be further
extended by two additional operators to the algebra of all (ground ring
valued) symplectomorphisms.  These additional operators are conserved, but
BRST invariant only weakly.

By standard topological arguments, the Lagrangian has its own class of
possible deformations that maintain topological invariance.

In \S{2.3} and \S{3.2} we have studied in some detail the topological
torus in the decompactified limit of $\im\tau_0\rightarrow\infty$, in which
just one dimension of the target remains compact, and the torus becomes an
infinite cylinder.  In this specific limit, spacetime algebra of topological
symmetry emerges, in which one of the non-exact symplectomorphisms mentioned
above plays the role of the BRST-like charge of topological symmetry in
spacetime.  As this charge is BRST invariant only weakly, spacetime
topological symmetry is hidden in the model.  It is interesting
to note that precisely this limit of the topological torus has been studied
in [\wtoporb], where some interesting conjectures about possible relations
to more familiar aspects of string theory, in particular to high energy
scattering, have been made.

In order to decide whether the topological theory is relevant to the
conventional phase of string theory in two dimensions, it is necessary to
find out which of these aspects persist after the topological torus is
coupled to topological gravity.  This issue has two sides -- first, we must
analyze the theory in a fixed gravitational background, which allows us
to extend the local results of this paper globally to surfaces of arbitrary
genus and solve the topological sigma model completely.   Next, we must
construct the full-fledged topological string theory on the torus.
Topological gravity becomes dynamical and contributes its own degrees of
freedom to the topological ground ring, as well as new generators to the
symmetry algebra.  Naively, one might expect one more spacetime dimension
to be generated by the Mumford-Morita cohomology classes $\sigma_m$
coming from the gravitational sector.

The model we have studied in this paper is a topological
string theory in two dimensions which, we believe, will be interesting in
itself.  Besides, we have found several analogies between this topological
theory and the conventional phase of string theory in two dimensions.  (Some
other analogies can be found in the light of the recent progress in two
dimensional string theory reported in [\wzwie].)  To what extent these
analogies are purely formal remains to be seen.  Clearly, much more must be
done before any detailed comparison of the topological theory studied in this
paper to the conventional phase of string theory in two dimensions will be
possible.

%%%%%%%%%%%%%%%%%%%%%%%%%%%%%%%%%%%%%%%%%%%%%%%%%%%%%%%%%%%%%%%%%%%%%%%%%%
\medskip
\ACK{It is a pleasure to thank Jeff Harvey, Emil Martinec and especially
Edward Witten for valuable discussions and useful comments on
preliminary versions of the manuscript.}

%%%%%%%%%%%%%%%%%%%%%%%%%%%%%%%%%%%%%%%%%%%%%%%%%%%%%%%%%%%%%%%%%%%%%%%%%%

\endpage

\refout
\end
\bye